%% file: alenex10.tex
\documentclass[11pt]{article}
\usepackage{fullpage,amsmath,amssymb,amscd,amsthm}
\usepackage{listings,algorithm,algpseudocode}
\usepackage{graphicx,subfigure}
\usepackage{enumerate}
\usepackage{url}
\usepackage{multirow}
\usepackage{array}
\usepackage{booktabs}

\graphicspath{{figs/},{../09-ANL-summer/},{../09-ANL-summer/figs/}}

\newcommand{\real}{\mathbb{R}}

\newcommand{\norm}[1]{\left\|#1\right\|}

\newcommand{\inprod}[2]{\left\langle #1,#2 \right\rangle}
\newcommand{\abs}[1]{\left|#1\right|}

\newcommand{\Romannumber}[1]{\uppercase\expandafter{\romannumeral #1}}

\newcommand{\ones}{\mathbf{1}}
\newcommand{\lap}{\mathcal{L}}
\newcommand{\Hh}{\mathcal{H}}
\newcommand{\Vh}{\mathcal{V}}
\newcommand{\Eh}{\mathcal{E}}

\DeclareMathOperator{\spn}{span}
\DeclareMathOperator{\range}{range}

\theoremstyle{definition} 
\theoremstyle{remark}     
\theoremstyle{remark}     
\theoremstyle{plain}      \newtheorem{theorem}{Theorem}
\theoremstyle{plain}      
\theoremstyle{plain}      
\theoremstyle{plain}      \newtheorem{corollary}[theorem]{Corollary}
\theoremstyle{plain}      

\renewcommand{\algorithmicrequire}{\textbf{Input:}}
\renewcommand{\algorithmicensure}{\textbf{Output:}}
\newcommand{\interval}[1]{{{\mathbb [} {#1} {\mathbb ]}}}

\begin{document}

%%%%%%%%%%%%%%%%%%%%%%%%%%%%%%%%%%%%%%%%%%%%%%%%%%%%%%%%%%%%
\title{\textsf{A Measure of the Connection Strengths
    between Graph Vertices with Applications}}
\author{
  Jie Chen%
  \thanks{Department of Computer Science and Engineering,
    University of Minnesota at Twin Cities, MN 55455.
    Email: \texttt{jchen@cs.umn.edu}. Work of this author is supported by NSF grant DMS-0810938, a University of Minnesota Doctoral Dissertation Fellowship, and the CSCAPES institute, a DOE project.}
  \and
  Ilya Safro%
  \thanks{Mathematics and Computer Science Division,
    Argonne National Laboratory, Argonne, IL 60439.
    Email: \texttt{safro@mcs.anl.gov}. This work was funded by the CSCAPES institute, a DOE project, and in part by DOE Contract DE-AC02-06CH11357.}
}
\date{}
\maketitle

%%%%%%%%%%%%%%%%%%%%%%%%%%%%%%%%%%%%%%%%%%%%%%%%%%%%%%%%%%%%
\begin{abstract}
We present a simple iterative strategy for measuring the connection strength between a pair of vertices in a graph. The method is attractive in that it has a linear complexity and can be easily parallelized. Based on an analysis of the convergence property, we propose a mutually reinforcing model to explain the intuition behind the strategy. The practical effectiveness of this measure is demonstrated through several combinatorial optimization problems on graphs and hypergraphs.

\if0
We present an analysis of a simple iterative method for measuring the connection strength between a pair of vertices in a graph and demonstrate its practical effectiveness on several combinatorial optimization problems on graphs and hypergraphs. Considering the influences of a neighborhood structure on an individual vertex, the strategy remedies the insufficiency of determining the couplings of the vertices solely on the edge weights information. The proposed method has a linear complexity and can be easily parallelized.
\fi

\end{abstract}

%%%%%%%%%%%%%%%%%%%%%%%%%%%%%%%%%%%%%%%%%%%%%%%%%%%%%%%%%%%%
\section{Introduction}
\par Measuring the connectivity between two vertices in a graph is one of the central questions in many theoretical and applied areas in computer science. A variety of methods exist for this purpose, such as shortest path length, number of paths between vertices, maximum flow, and minimum vertex/edge cut/separator. In this paper, we discuss a strategy that measures the connectivity between the vertices that are located not very far from each other. It is a simple iterative process that is based on the edge weights and that takes into account the neighborhood information of each vertex. We will analyze some properties of the strategy and demonstrate how the connectivity estimations can be used in practice to improve several well-known algorithms. In particular, the connectivity measure can be used in algorithms with greedy steps where it is critical to choose an appropriate edge that is the ``heaviest''.
\par Since the notion of connectivity is of practical significance, many algorithms have been developed to model it. In a random-walk approach \cite{rw-similar,social-network-book}, the average first-passage time/cost and average commute time were used. A similarity measure between nodes of a graph integrating indirect paths, based on the matrix-forest theorem, was proposed in \cite{vertex-proximity}. Approximation of a betweenness centrality in \cite{bader-betweenness} makes this computationally expensive concept feasible. A convergence of the compatible relaxation \cite{crs} was measured in Algebraic Multigrid (AMG) schemes \cite{vlsicad} in order to detect strong connections between fine and coarse points. A similarity method based on probabilistic interpretation of a diffusion was introduced in \cite{Nadler05diffusionmaps}. Our goal is to design a family of connectivity models and estimators that are fast and easy to implement and parallelize, and that can be applied on local parts of the data.

%\par The discussed measure of connectivity is called {\it algebraic distance}. Conceptually, a small distance means a strong connection. It was introduced in \cite{Ilya.relaxation} as a tool for the detection of strong couplings for the Algebraic Multigrid (AMG) based multilevel algorithms for combinatorial optimization problems. In a multilevel graph coarsening framework, one of the most vital concerns is how to choose vertices that are strongly coupled for merging (either in a strict or in a weighted sense), such that the coarse graph will faithfully represent the original one with respect to the given optimization problem \cite{vlsicad}.

%%%%%%%%%%%%%%%%%%%%%%%%%%%%%%%%%%%%%%%%%%%%%%%%%%%%%%%%%%%%
\section{Definitions and Notations}
\par Let $G=(V,E)$ denote a weighted simple connected graph, where $V=\{1,2,...,n\}$ is the set of nodes (vertices) and $E$ is the set of edges. Denote by $w_{ij}$ the non-negative weight of the undirected edge $ij$ between nodes $i$ and $j$; if $ij\notin E$, then $w_{ij}=0$. Let $W=\{w_{ij}\}$ be the weighted adjacency matrix of $G$. The graph Laplacian matrix is defined as $L=D-W$, where $D$ the diagonal matrix with diagonal elements $d_{ii}=\sum_jw_{ij}$. Correspondingly, the normalized Laplacian is $\lap=D^{-1/2}LD^{-1/2}$. Denote by $\delta_i$ the degree of vertex $i$.
\par The proposed strategy is to initially assign a random value $x_i$ to each vertex $i$ and to update the value $x_i$ one by one by a weighted combination of the value itself and the weighted average of $i$'s neighbors. Then, after a few iterations, the absolute difference between $x_i$ and $x_j$ is an indicator of the coupling between $i$ and $j$. This process is precisely stated in Algorithm~\ref{algo:alge.dist} in the vector form, where we use superscripts such as $^{(k)}$ and $^{(k-1)}$ to distinguish successive iterates and use subscripts to mean vector entries. We define the \emph{algebraic distance} (the discussed connectivity measure) between the vertex $i$ and $j$, at the $k$th iteration, to be
\begin{equation}\label{eqn:s_ijk}
s_{ij}^{(k)}:=\abs{x^{(k)}_i-x^{(k)}_j}.
\end{equation}
With $R$ initial vectors $x^{(0,r)}$, $r=1,\dots,R$, each vector is independently updated by using Algorithm~\ref{algo:alge.dist}, and the {\it extended $p$-normed algebraic distance} is defined as
\begin{equation}\label{eqn:ext_s_ijk}
\rho_{ij}^{(k)}:=\left( \sum_{r=1}^R \abs{x^{(k,r)}_i-x^{(k,r)}_j}^p \right)^{1/p}~,
\end{equation}
where the superscript $^{(k,r)}$ refers to the $k$th iteration on the $r$th initial random vector. For $p=\infty$, by convention, $\rho_{ij}^{(k)}= \max_{r=1}^R \abs{x^{(k,r)}_i-x^{(k,r)}_j}$.

\begin{algorithm}
\caption{Computing algebraic distances for graphs}
\label{algo:alge.dist}
\begin{algorithmic}[1]
\Require Parameter $\omega$, initial vector $x^{(0)}$
\For{$k=1,2,\dots$}
  \State $\tilde{x}_i^{(k)}\gets\sum_jw_{ij}x_j^{(k-1)}/\sum_jw_{ij}$, $\,\,\forall i$.
  \State $x^{(k)}\gets (1-\omega)x^{(k-1)}+\omega\tilde{x}^{(k)}$
\EndFor
\end{algorithmic}
\end{algorithm}
\par Conceptually, a {\it small} distance means a {\it strong} connection. The parameter $\omega$ is fixed to be $1/2$.

%%%%%%%%%%%%%%%%%%%%%%%%%%%%%%%%%%%%%%%%%%%%%%%%%%%%%%%%%%%%
\section{Historical Background and Motivation}
\par The algebraic distance is motivated by the Bootstrap AMG (BAMG) method~\cite{vlsicad} for solving a symmetric positive definite system $Ax=b$. In this method, a Gauss-Seidel (GS) process is run on the system $Ax=0$ in order to expose the slow-to-converge variables $x_i$ to allow a better interpolation of the low residual errors. Recently, the extended $\infty$-normed algebraic distance was used as a component of an AMG based coarsening scheme for graph linear ordering problems \cite{Ilya.relaxation}. In a multilevel graph coarsening framework, one of the most vital concerns is how to choose vertices that are strongly coupled for merging (either in a strict or in a weighted sense), such that the coarse graph will faithfully represent the original one with respect to the given optimization problem \cite{vlsicad}. Despite considerable empirical evidence of success in multilevel linear ordering algorithms, however, the concept of algebraic distance is still not well understood and has not been used widely in combinatorial optimization algorithms. This paper studies some properties of this relaxation process and interprets the algebraic distances under a mutually reinforcing environment model, where the neighborhood connectivity information governs the coupling of the vertices. In particular, two vertices are strongly connected if they are surrounded by similar neighborhoods. With this interpretation, the applications of this measure are no longer restricted to multilevel algorithms. Whenever the concept of vertex couplings is applicable, we can use the algebraic distances to measure the connectivity between vertices. We show a few such applications in this paper.

\par We note that an advantage of the proposed measure is its computational efficiency. As mentioned in the introduction, there exist other possible heuristics for measuring the connection strength between a pair of vertices, (e.g., the number of simple paths, the length of the shortest path, the commute time). However, these quantities are in general expensive to compute. For example, the problem of counting the number of simple paths connecting a pair of vertices is \#P-complete~\cite{num.paths.sharp.p.complete}, the Floyd-Warshall algorithm for computing all pairs shortest paths has an $O(n^3)$ time complexity, and to compute the commute times involves the pseudo-inverse of the Laplacian $L$. In contrast, since Algorithm \ref{algo:alge.dist} is essentially a Jacobi over-relaxation (JOR) process, its $k$ iterations take only $O(km)$ time, where $k$ is typically small, and $m$ is the number of edges in the graph. This time cost is a significant reduction, especially for sparse graphs. Further, the JOR iterations are easy to parallelize because unlike other iterative processes such as GS or SOR, the update of an entry $x_i$ does not require the most recent value of $x_{i-1}$. Thus, the proposed strategy has a strong potential for large-scale distributed computing.

%%%%%%%%%%%%%%%%%%%%%%%%%%%%%%%%%%%%%%%%%%%%%%%%%%%%%%%%%%%%
\section{Iterative Methods for Graph Laplacians}
\label{sec:classical.iterative}

Algorithm~\ref{algo:alge.dist}, on which the definition of algebraic distances is based, is essentially the JOR method for solving the linear system
\begin{equation}\label{eqn:Lx=0}
Lx=0.
\end{equation}
There being rich results for nonsingular systems, however, here the matrix $L$ is singular, and thus we need to first study the convergence properties for this particular system. In this section, we establish some general results for the convergence of several classical iterative methods (including JOR) for~\eqref{eqn:Lx=0}. The special case for JOR automatically applies.

Standard iterative methods by matrix splitting for solving a linear system can be written in a general form
\begin{equation}\label{eqn:x.iterate}
x^{(k+1)}=Hx^{(k)},\qquad k=0,1,2,\dots,
\end{equation}
where $H$ is the iteration matrix. Let the Laplacian $L=D-W_L-W_U$, where $W_L$ and $W_U$ are the strict lower and upper triangular parts of $W$, respectively. Then the iteration matrices for Gauss Seidel, Jacobi, SOR, and JOR are, respectively,
\begin{alignat*}{2}
H_{GS}&=(D-W_L)^{-1}W_U,
&\qquad H_{SOR}&=\left(D/\omega-W_L\right)^{-1}\left((1/\omega-1)D+W_U\right),\\
H_{JAC}&=D^{-1}(W_L+W_U),
&\qquad H_{JOR}&=\left(D/\omega\right)^{-1}\left((1/\omega-1)D+W_L+W_U\right).
\end{alignat*}
We will use the notation $H$ when the discussions are general or apply to all the iterative methods; we will add subscripts when an individual method is emphasized.

A matrix $A\in\real^{n\times n}$ is said to be \emph{convergent} if $\lim_{k\to\infty}A^k$ exists.\footnote{Note that some authors (e.g.,~\cite{horn.matrix.analysis}) use the term \emph{convergent} for a matrix $A$ where the limit $A^{k}$ is zero. However, the interesting case in this paper is that the limit is nonzero. Thus, we make a broader inclusion in the definition here.}

\begin{theorem}\label{thm:convergent}
The iteration matrices $H_{GS}$, $H_{SOR}$ \textup(with $0<\omega<2$\textup), and $H_{JOR}$ \textup(with $0<\omega<2/\rho(\lap)$\textup) are convergent, with spectral radius $\rho(H)=1$. The iteration matrix $H_{JAC}$ is convergent if and only if none of the connected component of the graph is bipartite.
\end{theorem}

We have the following result when $H$ is convergent.

\begin{theorem}\label{cor:iteration.convergence}
If the graph is connected, and the iteration matrix $H$ for the linear system~\eqref{eqn:Lx=0} is convergent, then the iterate $x^{(k)}$ converges to zero if the initial vector $x^{(0)}\in\range(I-H)$. Otherwise $x^{(k)}$ converges to a nonzero scalar multiple of $\ones$ \textup(a vector of all ones\textup).
\end{theorem}

\begin{corollary}\label{cor:alge.dist.convergence}
Under the conditions of Theorem~\ref{cor:iteration.convergence}, the quantity $s_{ij}^{(k)}$ defined in~\eqref{eqn:s_ijk} converges to zero for all $i$ and $j$.
\end{corollary}

To establish the rate of convergence, we make an additional mild assumption that $H$ is diagonalizable. Let $(\sigma_i,\phi_i)$ denote the eigen-pairs of $H$, where the eigenvalues are labeled in the order
\[
1=\sigma_1>|\sigma_2|\ge|\sigma_3|\ge\cdots\ge|\sigma_n|.
\]

\begin{corollary}\label{thm:convergence.rate.xk}
Under the conditions of Theorem~\ref{cor:iteration.convergence}, assume that $H$ is diagonalizable with eigen-pairs $(\sigma_i,\phi_i)$ labeled in nonincreasing order of the magnitudes of the eigenvalues. Then the iterate $x^{(k)}$ approaches the limit in the order $O(|\sigma_2|^k)$, and the quantity $s_{ij}^{(k)}$ defined in~\eqref{eqn:s_ijk} approaches zero in the same order.
\end{corollary}

The results of Corollaries~\ref{cor:alge.dist.convergence} and~\ref{thm:convergence.rate.xk} seem to suggest that the definition of the algebraic distance as a measure of the strength of connection is inappropriate. However, we are actually interested in comparing the relative magnitudes of $s_{ij}^{(k)}$ for different $(i,j)$ pairs. In other words, a concurrent scaling of the quantity $s_{ij}^{(k)}$ for all $i$ and $j$ will not compromise the measure. To this end, we consider the quantity
\begin{equation}\label{eqn:hat.s_ijk.eigen.basis}
\hat{s}_{ij}^{(k)}:=s_{ij}^{(k)}/\sigma_2^k.
\end{equation}
We have the following result.

\begin{theorem}\label{thm:convergence.algebraic.distance}
Under the conditions of Corollary~\ref{thm:convergence.rate.xk}, let the initial vector $x^{(0)}$ be expanded in the eigenbasis of $H$ as $x^{(0)}=a_1\phi_1+a_2\phi_2+\cdots+a_n\phi_n$.
\begin{enumerate}[\textup(i\textup)]
\item If $\sigma_2=\sigma_3=\cdots=\sigma_t$ and $|\sigma_t|>|\sigma_{t+1}|$ for some $t\ge2$, and if $a_2$, \ldots, $a_t$ are not all zero, then the quantity $\hat{s}_{ij}^{(k)}$ defined in~\eqref{eqn:hat.s_ijk.eigen.basis} approaches the limit $\abs{(e_i-e_j)^T\xi}$ in the order
$
O\left(\abs{\sigma_{t+1}/\sigma_t}^k\right)
$,
where $\xi$ is an eigenvector corresponding to the eigenvalue $\sigma_2$ \textup(with multiplicity $t-1$\textup).
\item If $|\sigma_2|=|\sigma_3|=\cdots=|\sigma_t|>|\sigma_{t+1}|$ for some $t\ge3$, where $\sigma_2,\dots,\sigma_t$ are not all the same, $a_2$, \ldots, $a_t$ are not all zero, and if there exists an integer $m$ such that $(\sigma_{\ell}/\sigma_2)^m=1$ for $\ell=3,\dots,t$, then the $p$-th subsequence $\{\hat{s}_{ij}^{(mk+p)}\}_{k=0,1,2,\dots}$ approaches the limit $\abs{(e_i-e_j)^T\eta_p}$ in the order
$
O\left(\abs{\sigma_{t+1}/\sigma_t}^{mk}\right)
$,
where
$
\eta_p=a_2\phi_2+a_3(\sigma_3/\sigma_2)^p\phi_3+\cdots+a_t(\sigma_t/\sigma_2)^p\phi_t
$
for $p=0,1,\dots,m-1$.
\end{enumerate}
\end{theorem}

%%%%%%%%%%%%%%%%%%%%%%%%%%%%%%%%%%%%%%%%%%%%%%%%%%%%%%%%%%%%
\section{Jacobi Over-Relaxations}
\label{sec:jor}
In this section, we discuss the implications of the theorems in the previous section for the JOR iterations. Specifically, the eigenvalues and vectors of the iteration matrix $H_{JOR}$ are closely related to those of the matrix pencil $(L,D)$. Because of the distributions of the eigenvalues, the convergence of Algorithm~\ref{algo:alge.dist} often is slow. Hence, we also study the behavior of the iterations at an early stage.

%%%%%%%%%%%%%%%%%%%%%%%%%%%%%%%%%%%%%%%%%%%%%%%%%%%%%%%%%%%%
\subsection{Algebraic Distances at the Limit}
An immediate result is that $H_{JOR}$ is diagonalizable and all the eigenvalues of $H_{JOR}$ are real, since
\[
H_{JOR}\,\,\phi_i=\sigma_i\phi_i \iff L\phi_i=\frac{1-\sigma_i}{\omega}D\phi_i.
\]
This equivalence implies that if $\mu_j$ is an eigenvalue of $(L,D)$, then $\mu_j=(1-\sigma_i)/\omega$ for some $i$. In general, we may not have $\mu_i=(1-\sigma_i)/\omega$ for all $i$, since the eigenvalues of $H_{JOR}$ are sorted in decreasing order of their magnitudes, whereas the eigenvalues of $(L,D)$ are sorted in their natural order. In particular, depending on the value of $\omega$, $\sigma_2$ can be either $1-\omega\mu_2$ or $1-\omega\mu_n$, and there are more possibilities for $\sigma_3$. Enumerating all the possible cases, we have the following theorem as a corollary of case (i) of Theorem~\ref{thm:convergence.algebraic.distance}.

\begin{theorem}\label{thm:jor}
Given a connected graph, let $(\mu_i,\hat{v}_i)$ be the eigen-pairs of the matrix pencil $(L,D)$, labeled in nondecreasing order of the eigenvalues, and assume that $\mu_2\ne\mu_3\ne\mu_{n-1}\ne\mu_n$. Unless $\omega=2/(\mu_2+\mu_n)$, the quantity $\hat{s}_{ij}^{(k)}$ defined in~\eqref{eqn:hat.s_ijk.eigen.basis} will always converge to a limit $|(e_i-e_j)^T\xi|$ in the order $O(\theta^k)$, for some $\xi$ and $0<\theta<1$. If $0<\omega<2/(\mu_2+\mu_n)$, then $\xi\in\spn\{\hat{v}_2\}$; otherwise \textup(if $2/(\mu_2+\mu_n)<\omega<2/\mu_n$\textup), $\xi\in\spn\{\hat{v}_n\}$.
\end{theorem}

\begin{figure}[htb]
\centering
\includegraphics{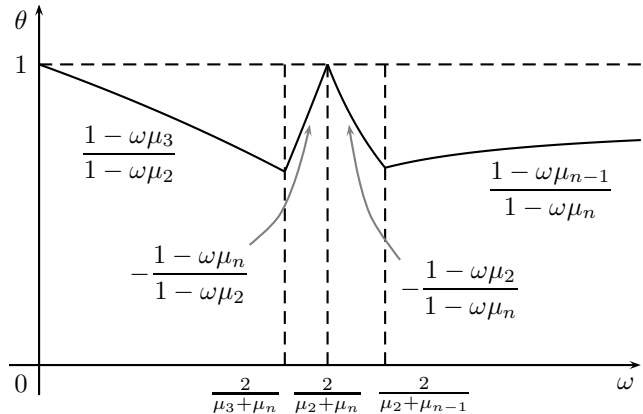}
\caption{The $\theta$ as a function of $\omega$. Note that the value $2/\mu_n$ can be less than, equal to, or greater than $2/(\mu_2+\mu_{n-1})$.}
\label{fig:theta}
\end{figure}

A graphical illustration of the dependence of $\theta$ on $\omega$ is shown in Figure~\ref{fig:theta}.

Theorem~\ref{thm:jor} shows two possible limits depending on the value of $\omega$. We can derive some upper/lower bounds for the cutting point $2/(\mu_2+\mu_n)$ and estimate which of the two cases in Theorem~\ref{thm:jor} is applied (will be given in full version of the paper). For example, if the graph is not complete, we have $2/(\mu_2+\mu_n)\ge2/3$, since in such a case $\mu_2\le1$. In practice, we deal with sparse graphs and we set $\omega=1/2$. Therefore, $\hat{s}_{ij}^{(k)}$ always converges to $|(e_i-e_j)^T\xi|$ with $\xi\in\spn\{\hat{v}_2\}$.

%%%%%%%%%%%%%%%%%%%%%%%%%%%%%%%%%%%%%%%%%%%%%%%%%%%%%%%%%%%%
\subsection{Algebraic Distances at Early Iterations}
Sometimes, even the optimal $\theta$ is so close to one that the theoretical convergence of $\hat{s}_{ij}^{(k)}$ is of little practical use---it takes an enormous number of steps before it gets close enough to the limit. (As observed for some real-life graphs, the smallest possible $\theta$ in Figure~\ref{fig:theta} can be as high as $0.999$.) However, an interesting phenomenon is that in practice $x^{(k)}$ soon becomes ``stable''; that is, the two iterates $x^{(k+1)}$ and $x^{(k)}$ are almost parallel even when $k$ is small.

\begin{theorem}\label{thm:iterate.angle.small}
Given a graph, let $(\mu_i,\hat{v}_i)$ be the eigen-pairs of the matrix pencil $(L,D)$, labeled in nondecreasing order of the eigenvalues. Denote $\hat{V}=[\hat{v}_1,\dots,\hat{v}_n]$. Let $x^{(0)}$ be the initial vector, and let $a=\hat{V}^{-1}x^{(0)}$ with $a_1\ne0$. If the following two conditions are satisfied:
\begin{subequations}
\begin{gather}
1-\omega\mu_n\ge0,\label{eqn:cond1}\\
f_k:=\frac{\alpha{r_k}^{2k}(1-{r_k})^2}{1+\alpha{r_k}^{2k}(1+{r_k})^2}\le\frac{1}{\kappa},\label{eqn:cond2}
\end{gather}
\end{subequations}
where $\alpha=\left(\sum_{i\ne1}a_i^2\right)/\left(4a_1^2\right)$, $r_k$ is the unique root of the equation
\[
2\alpha r^{2k+1}(1+r)=k-(k+1)r
\]
on the interval $[0,1]$, and $\kappa$ is the condition number of $D$, then
\begin{equation}\label{eqn:bound3}
1-\inprod{\frac{x^{(k)}}{\norm{x^{(k)}}}}{\frac{x^{(k+1)}}{\norm{x^{(k+1)}}}}^2\le\frac{4\kappa f_k}{(1+\kappa f_k)^2}.
\end{equation}
\end{theorem}

Note that when $\omega=1/2$, which is the relaxation parameter we use, \eqref{eqn:cond1} is satisfied, since the spectral radius of $(L,D)$, $\mu_n$, is less than or equal to $2$. Also note the condition number $\kappa$ of $D$. For many graphs arising from application areas such as VLSI design and finite element meshes, if the graph edges have a uniform weight equal to one, then $d_{ii}$ is the degree of a vertex. Thus, $\kappa$ will not be large. The bound~\eqref{eqn:bound3} serves as a reference for estimating how close to parallel two iterates are. In practice, for all the graphs we have experimented with, when $k$ is equal to $30$ or $50$, the quantity on the left-hand side of~\eqref{eqn:bound3} has dropped to the order of $10^{-4}$. Note that $\sin^2(\pi/180)=3.05\times10^{-4}$.

%%%%%%%%%%%%%%%%%%%%%%%%%%%%%%%%%%%%%%%%%%%%%%%%%%%%%%%%%%%%
\section{Mutually Reinforcing Model}\label{sec:interpretation}
The local structure of a graph and the edge weights are two factors that mutually govern the strength of the connection between a pair of vertices. We present here a model that incorporates both of the factors for quantitatively evaluating the vertex connectivity. Consider a mutually reinforcing environment, where entities are influenced by their neighbors. Intuitively, for an abstract property that is characteristic in such an environment, a part of the property value for an entity should be a weighted average of the influences from its neighbors in some way. Two entities are said to be close, or similar, if they are placed in two similar environments, or, consequently, their property values are close. If we consider the graph itself as an integral environment and vertices as individual entities each of which is surrounded by a neighborhood (the neighboring vertices), then two vertices are strongly coupled if they have similar values for an afore mentioned abstract property. Let each vertex $i$ be associated with a real number $x_i$. Except for a $\mu$ portion of itself, $i$ is influenced by its neighbors, which is quantitatively a weighted average:
\begin{equation}\label{eqn:mutual.reinforcement.informal}
x_i=\mu x_i + \sum_{j\sim i}p_{ij} x_j,
\end{equation}
where $j\sim i$ means $j$ is a neighbor of $i$. Here, the portion $0\le\mu\le1$ is an indicator of how strongly an environment will act on a vertex. When $\mu$ tends to zero, the neighborhood plays a major role, whereas when $\mu$ tends to one, a vertex is so stubborn that its neighbors cannot have a strong impact on it. The coefficient $\mu$ does not need to be explicitly specified; it is an internal property of the entire environment (i.e., the graph). For such a mutually reinforcing environment, a small $\mu$ is more desired. The weight of the influence by a neighbor $j$, $p_{ij}$, should be non-negative, and all the $p_{ij}$ related to the same $i$ should sum up to one. The weight $p_{ij}$ reflects how strong a neighbor can influence $i$, and therefore a natural choice is $p_{ij}=w_{ij}/\sum_{j}w_{ij}$. Thus,~\eqref{eqn:mutual.reinforcement.informal} is formally written in the following way
\begin{equation}\label{eqn:mutual.reinforcement.formal}
x_i=\mu x_i + \sum_j\frac{w_{ij}}{d_{ii}} x_j\qquad(0\le\mu\le1).
\end{equation}
Equivalently, in the matrix form, it is
\begin{equation}\label{eqn:mutual.reinforcement.formal.matrix.form}
x = \mu x + D^{-1}Wx.
\end{equation}
The coupling of two vertices $i$ and $j$ is measured by $\abs{x_i-x_j}$. A small value means a strong connection, which equivalently means that their neighborhoods have a similar influence on the two vertices.

\par From~\eqref{eqn:mutual.reinforcement.formal.matrix.form} we see that $x$ is an eigenvector of the matrix pencil $(L,D)$ and that $\mu$ is its corresponding eigenvalue. Unless the graph is complete, we have at least two sets of $x$ and $\mu$ that satisfy this system. In the first set, $\mu$ is zero, and $x$ is a nonzero scalar multiple of $\ones$. In this case, since $\mu=0$, the value of each vertex is entirely determined by its neighbors. This would have been the most desirable situation for a mutually reinforcing environment because it means that every entity is influenced only by its neighborhood. However, this situation leads to the result that every entity is the same ($x_i$ constant for all $i$), and therefore no discriminating power is presented. In the second set, $\mu$ is equal to $\mu_2$, the second smallest eigenvalue of $(L,D)$, and $x=\hat{v}_2$. When the graph is not complete, $\mu_2\le1$. Indeed, frequently $\mu_2$ is close to zero in practice. This is naturally a desirable solution for our problem: the neighborhood has a strong impact on a vertex, and vertices have different values such that the strengths of the connectivity for different vertex pairs can be distinguished.

We have proved that soon after the iterations start,  two successive iterates are getting close to parallel. In practice, we observe that when this situation happens, $x^{(k)}$ and $x^{(k+1)}$ form an acute angle and $\norm{H_{JOR}\,\,x^{(k)}}\approx(1-\omega\mu_2)\norm{x^{(k)}}$. (This makes sense because $\norm{H_{JOR}\,\,\hat{v}_2}=1-\omega\mu_2$.) Denote $\hat{x}^{(k)}=x^{(k)}/\norm{x^{(k)}}$. Then, we have
\[
\hat{x}^{(k)}\approx\mu_2 \hat{x}^{(k)}+D^{-1}W\hat{x}^{(k)}.
\]
This means that the (normalized) iterate $x^{(k)}$, when close to parallel to the next iterate, approximately satisfies the model~\eqref{eqn:mutual.reinforcement.formal}, with $\mu=\mu_2$. In other words, the algebraic distance $s_{ij}^{(k)}=\abs{x^{(k)}_i-x^{(k)}_j}$, computed from $x^{(k)}$, approximately measures the connection strength between $i$ and $j$ in our model.

We remark that for a small $k$, the iterate $x^{(k)}$ can be quite different from its limit $\hat{v}_2$, and for different initializations, $x^{(k)}$ will be different. However, they all satisfy or approximately satisfy the mutually reinforcing model~\eqref{eqn:mutual.reinforcement.formal}. This gives us the flexibility, yet not the arbitrariness, to estimate the connectivity for different vertex pairs. Readers may question why an iterate $x^{(k)}$ is preferred over the eigenvector $\hat{v}_2$ as the measure. A major reason is that the JOR method with a few number of iterations is computationally much less expensive than computing an eigenvector, even when the matrix $L$ (or pencil $(L,D)$) is sparse. Solving a large scale sparse eigenvalue problem for a symmetric matrix, say, using the Lanczos method~\cite{lanczos, saad.book.eigenvalue.problems, golub.book.matrix.computation}, involves frequent convergence tests, each of which needs to solve an eigenvalue subproblem for a tridiagonal matrix. On the other hand, inside each JOR iteration is nothing but weighted averages. Therefore, the JOR process is particularly inexpensive compared with computing an eigenvector. Besides, the simplicity of Algorithm~\ref{algo:alge.dist} makes it particularly attractive, and thus it is advocated as \emph{the} algorithm for the proposed measure in this paper.

\input{paper-applications}

%%%%%%%%%%%%%%%%%%%%%%%%%%%%%%%%%%%%%%%%%%%%%%%%%%%%%%%%%%%%
\section{Conclusion}
In this paper, we proposed a simple iterative algorithm for measuring the connectivity between graph vertices. We presented a convergence analysis of the algorithm, and interpreted the proposed measure by using a mutually reinforcing model. Empirical results show that the proposed measure is effective in several combinatorial optimization problems. 

%%%%%%%%%%%%%%%%%%%%%%%%%%%%%%%%%%%%%%%%%%%%%%%%%%%%%%%%%%%%
\bibliographystyle{plain}

\bibliography{reference,ilya-biblio}
\hspace*{1.5in}{\scriptsize\framebox{\parbox{2.4in}{
The submitted manuscript has been created in part by UChicago Argonne, LLC, Operator of Argonne National Laboratory (``Argonne'').  Argonne, a U.S. Department of Energy Office of Science laboratory, is operated under Contract No. DE-AC02-06CH11357.  The U.S. Government retains for itself, and others acting on its behalf, a paid-up nonexclusive, irrevocable worldwide license in said article to reproduce, prepare derivative works, distribute copies to the public, and perform publicly and display publicly, by or on behalf of the Government.
}}}

\end{document}

%% file: paper-applications.tex
%%%%%%%%%%%%%%%%%%%%%%%%%%%%%%%%%%%%%%%%%%%%%%%%%%%%%%%%%%%%
\section{Applications}\label{sec:app}
\par In this section, we demonstrate how the algebraic distance can be used in practice. For this purpose, we have chosen four problems: maximum weighted matching, maximum independent set and the minimum $\tau$-partitioning of graphs and hypergraphs. In all these cases, fast existing baseline algorithms were modified by taking into account the algebraic distance instead of the original graph edge weights. The experimental graphs were of different sizes ($|E|$ was between $10^3$ and $10^7$) and have been selected from the UFL database of real-life matrices \cite{davis}. Because of page limitation, we present only two applications here. The other two applications will be included in the full paper.

\subsection{Maximum Weighted Matching}
\par A matching, $M$, of $G$ is a subset of $E$ such that no vertex in $V$ is incident to more than one edge in $M$. A matching $M$ is said to be {\it maximum} if, for any other matching $M'$, $|M|\geq |M'|$. Similarly, a matching $M$ is said to be {\it maximum weighted} if, for any other matching $M'$, $w(M)\geq w(M')$, where $w(M) = \sum_{ij\in M} w_{ij}$.
\par In many practical applications two well-known 2-approximations are used. One is a textbook greedy algorithm \cite{CormenLeisersonRivest1993} for maximum weighted matching; the other is its improved version which is based on the path growing principle. Both algorithms are presented in \cite{drake-matching}.
\par Based on these algorithms and the algebraic distances, we designed two heuristics for the maximum weighted matching. In both algorithms, there exists a greedy step in which the next heaviest edge has to be chosen. The criterion for choosing an edge was changed according to the following heuristic observation: a better matching can be found in a dense graph with less effort than in a sparse graph. According to this observation, we give preference to matching two nodes that are not connected well with other nodes from their neighborhood. The preprocessing for greedy algorithms is presented in Algorithm \ref{alg:matchingprep}.
\begin{algorithm}
\caption{Preprocessing for greedy algorithm for maximum matching}
\label{alg:matchingprep}
\begin{algorithmic}[1]
\Require Graph $G$\\
For all edges $ij\in E$ calculate $\rho_{ij}^{(k)}$ for some $k,~R$ and $p$\\
For all nodes $i\in V$ define $a_i=\sum_{ij\in E}1/\rho_{ij}^{(k)}$\\
For all edges $ij\in E$ define $s'_{ij}=a_i / \delta_i + a_j / \delta_j$
\end{algorithmic}
\end{algorithm}
\par The resulting values $s'_{ij}$ are used in greedy choice steps (and sorting if applicable) instead of graph weights. The experimental results of the comparison are presented in Figure \ref{fig:applic}a as ratios between the textbook greedy matching {\it with} preprocessing and the same algorithm {\it without} preprocessing. Almost identical results were obtained by improving a greedy path growing algorithm from \cite{drake-matching}. These particular results were obtained with $k=20$, $R=10$, and $p=\infty$. However, results of almost the same quality have been obtained with many different combinations of $R\geq 5$, $10 \leq k \leq 100$, and $p=1,2$.
\subsection{Hypergraph Partitioning}
\par Extending the algebraic distance for hypergraphs is the next step in our future research. Here we present some preliminary results of defining the algebraic distances on hypergraphs and experimenting with them on the hypergraph partitioning problem.
\par We define a hypergraph $\Hh$ as a pair $\Hh=(\Vh,\Eh)$, where $\Vh$ is a set of nodes and $\Eh$ is a set of hyperedges. Each $h\in\Hh$ is a subset of $\Vh$. A hypergraph $\tau$-partitioning is a well-known NP-hard problem (see \cite{gajost76} for its graph version). The goal of the problem is to find a partitioning of $\Vh$ into a family of $\tau$ disjoint nonempty subsets $(\pi_p)_{1\leq p \leq \tau}$, while enforcing the following:
\begin{equation}
\begin{split}
   \text{ minimize } & \mathop{\sum_{h\in \Eh \text{ s.t. } \exists i,j\in h \text{ and}}}_{i\in \pi_p \Rightarrow j \not\in \pi_p}w_{h} \\
   \text{ such that }& \forall p \in \interval{1,\tau}, |\pi_p| \leq (1 + \alpha) \cdot \frac{|V|}{\tau}~,
\end{split}
\end{equation}
where $\alpha$ is a given {\it imbalance factor}. In this paper, we refer to the problem with $\tau=2$.
\par HMetis2 \cite{metis} is one of the fastest and most successful modern solvers for partitioning problems. We use it as a black-box solver and define its extension HMetis2+ in Algorithm \ref{alg:adhyper}. First, we extend the algebraic distance for hypergraphs using its bipartite model. In particular, we create $G=(V,E)$ with $V=\Vh \bigcup \Eh$ and $ij\in E$ if $i\in \Vh$ appears in edge $j\in \Eh$. The edge weights are preserved with no change. Second, HMetis2 is applied on hypergraph with new hyperedge weights, namely inverse ({\it small} algebraic distance replaces {\it heavy} edge weight) algebraic distances.
\begin{algorithm}
\caption{HMetis2+ : First eight lines---algebraic distance related preprocessing}
\label{alg:adhyper}
\begin{algorithmic}[1]
\Require Hypergraph $\Hh$, $k=20$, $R=10$\\
$G=(V,E)~\leftarrow$ bipartite graph model\\
Create $R$ initial vectors $x^{(0,r)}$
\For{$r=1,2,\dots,R$}
\For{$m=1,2,\dots,k$}
  \State $x_i^{(m,r)}\gets\sum_jw_{ij}x_j^{(m-1,r)}/\sum_jw_{ij}$, $\,\,\forall i$.
\EndFor
\EndFor
\State \Return modified edge weights $s_{h}^{(k)}\gets \sum_r\max_{i,j\in h} \abs{x_i^{(k,r)}-x_j^{(k,r)}}$, $\,\,\forall h\in \Eh$.
\end{algorithmic}
\begin{algorithmic}[1]
\Require Hypergraph $\Hh$ with hyperedge weights $1/s^{(k)}_h$\\
$C~ \leftarrow$ hyperedge cut obtained by HMetis2 on $\Hh$ with the modified edge weights\\
\Return cost of $C$ with original edge weights
\end{algorithmic}
\end{algorithm}
%\par A generalization of the graph partitioning problem, a hypergraph partitioning problem, is of practical significance in many applications \cite{ZoltanParHyp06ipdps}. In the hypergraph partitioning problem, a hyperedge belongs to the cut if not all its nodes belong to the same part. Similarly to the experiments with the graph 2-partitioning problem, we experimented with HMetis2 for obtaining 2-partitioning for hypergraphs. In these experiments, we designed the same algebraic distance based algorithm HMetis2+ (see Algorithm \ref{algo:hmetis}) with a preprocessing for calculating algebraic distances for hypergraphs.
\par
\begin{figure}
\centering
\subfigure[]{\includegraphics[width=3in]{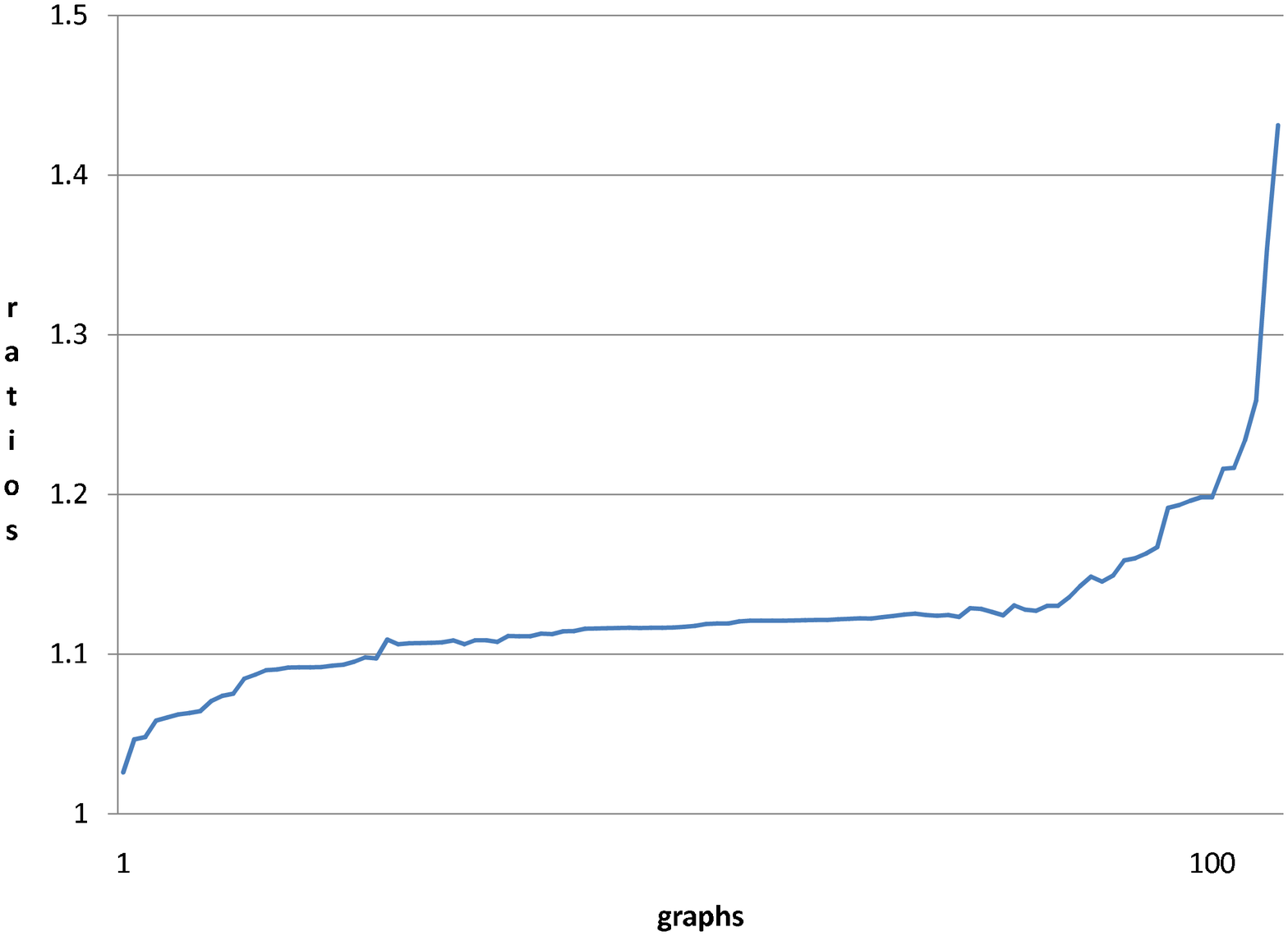}}
\subfigure[]{\includegraphics[width=3in]{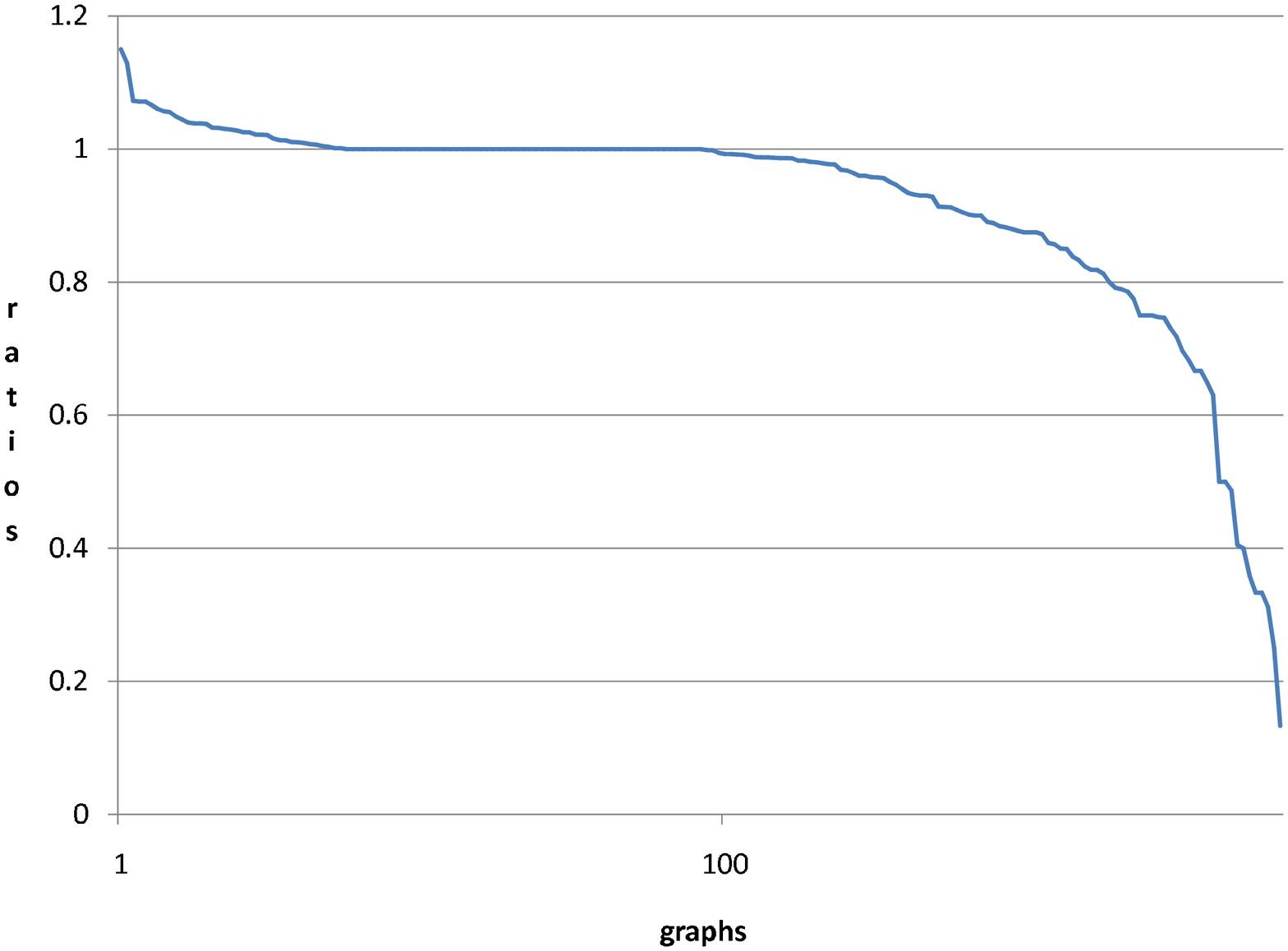}}
\caption{{\bf (a)} Comparison of greedy algorithms for matching with and without algebraic distance preprocessing. Each point corresponds to the average of ratios between matching sizes costs produced by the greedy algorithm with preprocessing and the same algorithm without preprocessing for one graph. The average was calculated over 20 different executions with different random initial vectors. The total number of graphs is 100. {\bf(b)} Comparison of HMetis2 and HMetis2+ for hypergraph 2-partitioning. Each point corresponds to the average of ratios between cut costs produced by HMetis2 and HMetis2+ for one graph. The average was calculated over 20 different executions with different random seeds. The total number of graphs is 200.}\label{fig:applic}
\end{figure}

 The numerical results of comparing HMetis2 and HMetis2+ are presented in Figure \ref{fig:applic}b. Since HMetis2 is a multilevel algorithm, a more correct way to apply the algebraic distances is to use them at all levels as it was demonstrated in \cite{Ilya.relaxation}. Even in these preliminary results, however, one can easily see that a significant improvement can be obtained by reinforcing a black-box algorithm with algebraic distances.

\if0
\begin{figure}[ht]
\centering
\includegraphics[width=.9\linewidth]{hypergraph-part}
\caption{Comparison of HMetis2 and HMetis2+ for hypergraph 2-partitioning. Each point corresponds to the average of ratios between cut costs produced by HMetis2 and HMetis2+ for one graph. The average was calculated over 20 different executions with different random seeds. The total number of graphs is 200.}
\label{fig:hpartres}
\end{figure}
\fi

%%%%%%%%%%%%%%%%%%%%%%%%%%%%%%%%%%%%%%%%%%%%%%%%%%%%%%%%%%%%